\documentclass{PoS}

\title{New approach to the Parton Distribution Functions: Self-Organizing 
Maps}

\ShortTitle{New approach to the Parton Distribution Functions: 
Self-Organizing Maps}

\author{\speaker{Heli Honkanen}\thanks{We thank our computer science 
collaborators D.~Brogan, J.~Carnahan, Y.~Loitiere and  P.~R.~Reynolds. HH also gratefully
acknowledges the travel grant from Gary McCartor Fund and the ``research 
equipment'' from Sheila  McCartor. 
This work was financially supported by the US
National Science Foundation grant no.0426971. 
HH was also supported by the U.S. Department of Energy, grant no.
DE-FG02-87ER40371. SL is supported by the U.S. Department of Energy, grant no.
DE-FG02-01ER41200.}\\
            Department of Physics and Astronomy, 
            Iowa State University, Ames, IA 50011, USA\\
        E-mail: \email{heli@iastate.edu}}

\author{Simonetta Liuti\\
        Department of Physics, University of Virginia,
           Charlottesville, VA 22904-4714, USA\\
        E-mail: \email{sl4y@virginia.edu}}

\abstract{ We propose a Parton Distribution Function (PDF) fitting
technique which is based on an
interactive neural network algorithm using Self-Organizing Maps (SOMs).
SOMs are visualization algorithms based on competitive learning
among spatially-ordered neurons. Our SOMs are trained with
stochastically generated PDF samples. On every optimization iteration
the PDFs are clustered on the SOM according to a user-defined feature and
the most promising candidates are selected as a seed for the subsequent
iteration.
Our main goal is thus to provide a fitting procedure that, at variance with
the global analyses and
standard neural network approaches, allows for an increased control of
the systematic bias by enabling  user interaction in the various stages of
the fitting process.}

\FullConference{LIGHT CONE 2008 Relativistic Nuclear and Particle Physics\\
		 July 7-11  2008\\
		 Mulhouse, France}

\begin{document}

\section{Introduction}
The cross sections for a number of hadronic reactions can be computed
using perturbative Quantum Chromodynamics (pQCD) 
convoluting the perturbatively 
calculable hard scattering coefficients  with the non perturbative 
Parton Distribution Functions (PDFs), that parametrize the large distance 
hadronic structure.
The accuracy with which the theoretical predictions for observables
of such reactions can be compared against the  high precision experimental
data thus depends, not only on the accuracy of the hard scattering part 
calculations, but also on the accuracy with which the PDFs are known.

Currently, the established method to obtain the 
PDFs, used by the major  PDF collaborations 
(CTEQ \cite{Nadolsky:2008zw} and references within, MRST
\cite{Martin:2001es}, 
Alekhin \cite{Alekhin:2002fv}, Zeus \cite{Chekanov:2005nn}
and H1 \cite{Adloff:2003uh}),
is the global analysis supplemented with an error estimation using
some kind of variant of
 the Hessian method (see e.g. \cite{Pumplin:2001ct}
for details).
This powerful combination allows for both extrapolation 
outside the kinematical range of the data and extension to multivariable cases,
 such as nuclear PDFs. 
However, there are uncertainties related to the method itself, that are 
difficult to quantify, but may turn out to have a large effect.
The differences between the current global PDF sets indeed
tend to be larger than the estimated uncertainties \cite{Pumplin:2005yfa},
and these differences again translate to the predictions for the LHC 
observables, such as Higgs  \cite{Djouadi:2003jg} or $W^\pm$ and $Z$ production
cross sections  \cite{Nadolsky:2008zw}.
 For details of PDF uncertainty studies  see e.g. 
Refs.~\cite{Martin:2003sk}.

Another approach to the PDF fitting has recently been proposed by the
NNPDF collaboration \cite{Ubiali:2008uk},
who have replaced typical functional form ansatze used in global analyses
with  more complex  standard neural network (NN) solutions, and the Hessian 
method with Monte Carlo (MC) sampling of the data. 
The NNPDF method circumvents many of the problems global analyses suffer, such
as bias resulting from fixing a functional form and selecting a suitable 
tolerance $\Delta \chi^2$ needed in Hessian method,
and it relies on genetic algorithm (GA) which 
works on a population of solutions for each MC replica of the data, thus having
a lowered possibility of getting fixed in local minima.
The estimated uncertainties for NNPDF fits are larger than those of global
fits, possibly indicating that the global fit uncertainties may have been
underestimated. 
The complexity  of NN results, however,
 may also induce problems, especially when used in
a purely automated fitting procedure. Since the effect of modifying individual
NN parameters is unknown, the result may exhibit strange or unwanted behaviour
in the extrapolation region, or in between the data points if the data is 
sparse. 
Implementation of information not given directly by the data, such
as nonperturbative models, lattice calculations or  knowledge from prior work
in general, is also difficult in this approach. 

The new PDF fitting method we have recently proposed in Ref.~\cite{Carnahan:2008mb}
relies on the use of 
Self-Organizing Maps (SOMs), a subtype of neural network. The
idea of our method is to create means for introducing
``Researcher Insight'' instead of ``Theoretical bias'' by giving up
a fully automated fitting procedure, and eventually 
to develop an interactive
fitting program which would allow us 
to combine the best features of both the global analysis and 
the NNPDF approach.

\section{Self-Organizing Maps}
The SOM \cite{Teuvo}
 is a visualization algorithm which
attempts to represent all the available observations with optimal accuracy 
using a restricted set of models.
SOM consists of nodes, map cells, which are all assigned spatial 
coordinates, and  the topology of the map is determined by a chosen distance 
metric $M_{\mathrm{map}}$. Each cell $i$ contains  a  map vector $V_i$,
that is isomorphic to the 
data samples used for training of the neural network. 
For a simple 2-dimensional rectangular lattice, our choice for the SOM shape, 
a natural choice for the topology is  
$L_1(x,y)=\sum_{i=1}^2\vert x_i-y_i\vert$.

\begin{figure}[h]
\begin{center}
\vspace{-1.2cm}
\includegraphics[width=15cm]{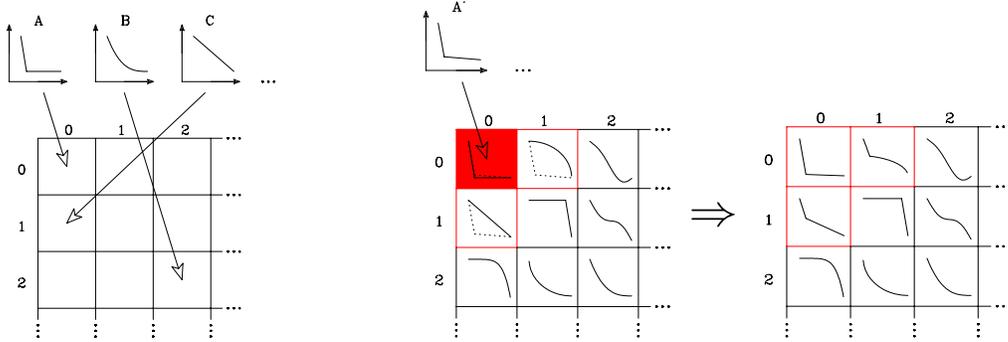} \hspace{0.0cm}
\vspace{-1.2cm}
\caption[a]{\protect \small {\bf Left:} SOM initialization,
{\bf Right:} SOM training.}
\label{train2.eps}
\end{center}
\end{figure}

The implementation
 of SOMs proceeds in three stages: 1) initialization of the SOM 
(see Fig.~\ref{train2.eps}),
2) training of the SOM (Fig.~\ref{train2.eps})
 and 3) associating the data samples with a
trained map, i.e. clustering. For the details of the SOM implementation,
see \cite{Carnahan:2008mb}.
\vspace{-0.0cm}

In the end of the training stage, cells  that are topologically close to 
each other have map vectors which are most similar to each other (according to
a chosen similarity metric $M_{\rm data}$) compared to
all the other map vectors.
In the matching phase the actual data is matched against the 
map vectors of the trained map, and thus get distributed on the map according 
to the feature that was used as the similarity criterion. Clusters now
emerge as a result of {\it unsupervised learning}. 
This local similarity property is the feature that makes SOM suitable for 
visualization purposes, 
thus facilitating user interaction with the data. Since each map vector now
represent a class of similar objects, the SOM
is an ideal tool to
visualize high-dimensional data, by projecting it onto a low-dimensional map
clustered according to some desired similar feature.

In our work we used the so-called batch-version of the training, in which
all the training data samples are matched against the map vectors before the
training begins. The map vectors 
 are then averaged with all the training  samples within the neighbourhood 
radius simultaneously.
 The procedure is repeated  $N_{\mathrm{step}}$ (free parameter to 
choose) times such that in
every training step the {\it same} set of training data samples is associated
with the evolving map
The benefit of the batch training compared to the incremental training,
shown in Fig.~\ref{train2.eps}, is that the training is independent of 
the order in which the training samples are introduced on the map.

\section{ENVPDF algorithm}
The aim of our approach is to both i) to be able to study the 
properties of the PDFs in a model independent way and yet ii) to be able to 
implement knowledge from the prior works on PDFs, and ultimately iii)
to be able to guide the fitting procedure interactively with the help of 
the SOM properties.

To accomplish this, we choose, at variance with the ``conventional'' PDFs
sets or NNPDFs, to give up the functional form for the PDFs 
and rather to rely on purely stochastical methods in 
generating the initial and training PDF samples. Our choice is
a GA-type analysis, in which our parameters are the values of PDFs at the 
initial scale 
for each flavour at each value of $x$ where the experimental data exist.
To obtain control over the shape of the PDFs  we use some of the
existing distributions 
to establish an initial range, or {\em envelope}, within which we sample the 
candidate PDF values. 

For now we concentrate on
DIS structure function data from H1 \cite{Adloff:2000qk}, 
BCDMS \cite{Benvenuti:1989rh} and Zeus \cite{Chekanov:2001qu},
which we use without additional kinematical cuts or normalization factors.
The parameters for the DGLAP scale evolution were chosen to be those of
 CTEQ6 (CTEQ6L1 for lowest order (LO)) \cite{Pumplin:2002vw:cteq6}, 
the initial scale being 
$Q_0=1.3$ GeV. In next-to-leading order
(NLO) case the evolution code was taken from \cite{QCDNUM} 
(QCDNUM17 beta release).

We use  CTEQ6 \cite{Pumplin:2002vw:cteq6},
CTEQ5 \cite{Lai:1999wy:cteq5},
MRST02 \cite{Martin:2001es,Martin:2002dr}, Alekhin \cite{Alekhin:2002fv} and 
GRV98 \cite{Gluck:1998xa} PDF sets
as our {\it init} PDFs. We construct our initial PDF generator first
to, for each flavour separately, 
select randomly  either the range  $[0.5,1]$, 
$[1.0,1.5]$ or $[0.75,1.25]$ times any of the init PDF set.
Next the initial generators  generate values for  
each $x_{\rm data}$ (
To ensure a reasonable large-$x$ behaviour for the PDFs, we also generate
with the same method
values for them in a few $x$-points outside the range of the experimental
data. For simplicity we also require the gluons to be positive in NLO.)  
using uniform, instead of Gaussian, 
distribution around the init PDFs, thus reducing direct bias from them.
Gaussian smoothing is applied to the 
resulting set of points, and  the flavours 
combined to form a PDF set such that the curves are linearly 
interpolated from the discrete set of generated points, and scaled to
conserve momentum, baryon number and charge.
 In this study we accept
the $<$few\% normalization error which results from the fact that our 
x-range is not $x=[0,1]$, but  $x=[{\rm min}(x_{\rm data}),1]$.
We call these type of PDF sets {\it database} PDFs.

For a $N\times N$ SOM we choose the size of the database to be $4N^2$.
We randomly initialize the map with  $N$ database PDFs sets,
such that each map vector $V_i$
consists of the PDF set itself, and
of the observables $F_2^p(x,Q_0^2)$  derived from it, and
train the map with  $N_{\mathrm{step}}$ batch-training steps.
In order to obtain a reasonable selection of PDFs to start with, we reject 
candidates which have $\chi^2/N>10$.
We choose the similarity criterion to be the 
similarity of observables $F_2^p(x,Q^2)$ with
$M_{\mathrm{data}}=L_1$. The similarity is tested at every
$x_{\rm data}$-values both at the initial scale and at all the evolved 
scales where experimental data exist.
On every training step, after the matching, all 
the observables (PDFs) of the map vectors get averaged with the observables 
(PDFs, flavor by flavor) matched within the neighbourhood.
The resulting new averaged map
vector PDFs are rescaled again to obey the sumrules. We call
these type of PDF sets {\it map} PDFs. 
The map  PDFs  are evolved and the 
observables at every experimental data scale are computed
and compared for similarity with  the observables from the 
training PDFs. After the training we have a map with  $N$ map PDFs and the 
same $4N^2$ database PDF sets we used to train the map. 
This is the end of the first optimization {\it iteration}.

During the later iterations we proceed as follows:
At the end of each iteration we pick from the trained SOM $2N$ 
best PDFs as the init PDFs.
These init PDFs are introduced into the training set alongside with the 
database PDFs, which are now constructed using each of the init PDFs 
{\it in turn} as a center for a Gaussian random number generator, which 
assigns for {\it all} the flavours for each $x$ a value around that {\it same}
 init PDF
such that $1-\sigma$ of the generator is given by the spread of the best PDFs 
in the topologically nearest neighbouring cells. 
The object of these generators is thus to refine a good candidate PDF found 
in the previous iteration by jittering its values within a  range
determined by the shape of other good candidate PDFs from the previous 
iteration. The generated PDFs are then  smoothed and
scaled to obey the sumrules. Sets with  $\chi^2/N>10$ are always rejected.
It is important to preserve
the variety of the PDF shapes on the map, so
we also keep $N_{\rm orig}$ copies of the first iteration 
generators in our generator mix. Since the best PDF candidates from the 
previous iteration are matched on this new map as an unmodified
init PDF, it is guaranteed that the  $\chi^2/N$ as a function of the
iteration either decreases or remains the same.  
We keep repeating the 
iterations until the $\chi^2/N$ saturates. 

The  best $\chi^2/N$ values of the original
init PDFs\footnote{These are the  $\chi^2/N$ for the initial scale
PDF sets taken from the quoted parametrizations and
evolved with CTEQ6 DGLAP settings, no kinematical cuts
or normalization factors for
the experimental data were imposed. 
We do not claim these 
values to describe the quality of the quoted PDF sets.}
are  1.67 for LO (CTEQ6) and 1.89 for NLO (MRST02), and
Table~\ref{envpdftab} lists results from a variety of ENVPDF runs. The results
do not seem to be very sensitive to the number of SOM training steps,
$N_{\rm step}$, but are highly sensitive to the number of first iteration 
generators used in subsequent iterations. Although the generators can 
now in principle produce an infinite number of different PDFs, the algorithm 
would not be able
to radically change the shape of the database PDFs without introducing
a random element on the map. Setting $N_{\rm orig}> 0$ provides,
through map PDFs, that element, and keeps the algorithm from
getting fixed to a local minimum. 

\begin{table}[h]
\center
\begin{tabular}{|c|c|c|c|c|c|}
\hline
SOM & $N_{\rm step}$ & $N_{\rm orig}$
& LO $\chi^2/N$ & NLO $\chi^2/N$\\ \hline
5x5 & 5  & 2 & 1.04 & 1.08  \\ \hline
5x5 & 10  & 2 & 1.10 & - \\ \hline
5x5 & 20  & 2 & 1.10 & - \\ \hline
5x5 & 30  & 2 & 1.10 & - \\ \hline
5x5 & 40  & 2 & 1.08 & - \\ \hline
5x5 & 5  & 0 & 1.41 & - \\ \hline
15x15 & 5  & 6 & 1.00 & 1.07 \\ \hline
\end{tabular}
\caption{$\chi^2/N$ for variety of ENVPDF runs against all the datasets 
(H1, ZEUS, BCDMS, N=709).}
\label{envpdftab}
\end{table}

\begin{figure}[h]
\begin{center}
\vspace{-1.2cm}
\hspace*{-1.0cm}\includegraphics[width=12cm]{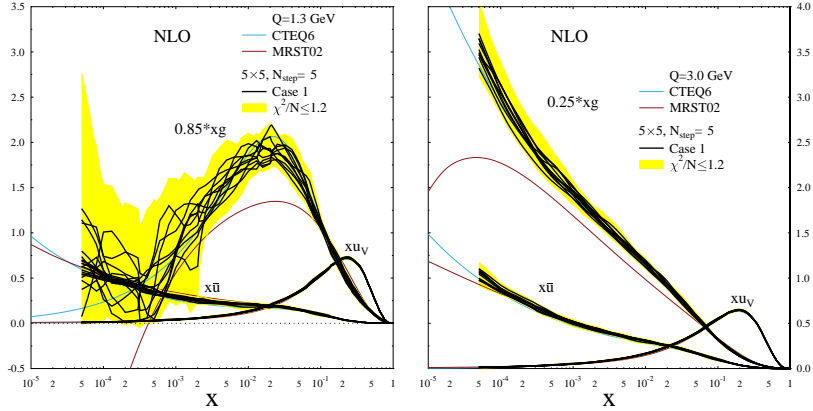} \hspace{0.0cm}
\vspace{-0.5cm}
\caption[a]{\protect \small  
NLO ENVPDF best results and the  $\chi^2/N\le 1.2$
spreads of results from 10 separate runs.}
\label{envpdf_jakaumat_case1_best_nlo.eps}
\end{center}
\end{figure}

Due to the stochastical nature of the ENVPDF algorithm, we may
well study the combined results from several separate runs.
It is especially important to verify the stability of our results, to show 
that the results are indeed reproducible instead of lucky coincidences.
Left panel of Fig.~\ref{envpdf_jakaumat_case1_best_nlo.eps} presents the best 
NLO results, and 
the combined $\chi^2/N\le 1.2$ spreads of the PDFs from any iteration, for
10 repeated $5\times 5$, $N_{\rm step}=5$ runs at the initial scale. 
The average $\chi^2/N$ and the standard deviation $\sigma$ for these runs
are 1.122 and 0.029, corresponding
to $\Delta\chi^2\sim 20$. 
The right panel of
the same  Fig.~\ref{envpdf_jakaumat_case1_best_nlo.eps} 
shows the 10 best result curves
and the  $\chi^2/N\le 1.2$ spreads evolved 
up to $Q=3.0$ GeV.
 Since we have only used DIS data in this study, we are 
only able to explore the small-$x$ uncertainty for now, and
expectedly, the small-$x$ gluons obtain the 
largest uncertainty for all the cases we studied. 

Clearly the seemingly large difference between the small-$x$ gluon results
at the initial scale is not statistically significant, but gets smoothed out
during the course of  the QCD evolution.
The evolved curves also preserve the
initially set baryon number scaling within
$~0.5\%$ and momentum sumrule  within $~1.5\%$ accuracy.
Thus the initial
scale wiggliness of the PDFs is mainly only a residual effect from our method 
of generating  them and  not linked to the overtraining
of the SOM.

Therefore our simple method of
producing the candidate PDFs by jittering random numbers inside a predetermined
envelope is surprisingly stable when used together with a complicated PDF 
processing that SOMs provide.
Remarkably then, even a single SOM run can provide a quick uncertainty estimate
for a chosen  $\Delta\chi^2$ without performing a separate error analysis.

\vspace{-0.0cm}

\section{Future of the SOMPDFs}
So far we have shown a relatively straightforward method of obtaining 
stochastically generated, parameter-free, PDFs, with an uncertainty estimate
for a desired $\Delta\chi^2$.
However, the proposed method can be extended much further than that.
What ultimately sets the SOM method apart from the standard global analyses or
NNPDF method  are the clustering and 
visualization possibilities that it offers. 
Instead of setting $M_{\mathrm{data}}=L_1$ and clustering according to the
similarity of the observables, it is possible to set the clustering criteria
to be anything that can be mathematically quantified, e.g. the shape of the
gluons or the large-$x$ behaviour of the PDFs. 
The desired feature of the PDFs can then be projected out from the SOM.
Moreover, by combining the method with an interactive graphic user 
interface (GUI), it
would be possible to change and control the shape and the width of the 
envelope as the
minimization proceeds, to guide the process by applying researcher insight at 
various stages of the process, and the uncertainty band produced by the
SOM could further
help the user to make decisions about the next steps of the minimization.
With GUI it would be e.g. possible to set the generators to
 sample a vector
consisting of PDF parameters, instead of values of PDFs in each value of
$x$ of the data. That would lead to smooth, continuous type of solutions, 
either along the lines of global analyses, or NNPDFs using $N$ SOMs
for $N$ Monte-Carlo sampled replicas of the data. 
For such a method, all the existing error estimates, 
besides an uncertainty band produced by the map,
would be applicable as well.
Since the solution would be required to stay within an envelope of selected
width and shape, no restrictions for the parameters themselves
would be required, and it would be possible to e.g. to
constrain the extrapolation of the NN 
generated PDFs outside the $x$-range of the data without explicitly
introducing terms to ensure the correct small- and large-$x$ behaviour
as in NNPDF method. 
The selection of the best PDF candidates for the subsequent iteration could 
then be made based on the user's preferences instead of solely based on the
 $\chi^2/N$.
That kind of method in turn could be extended to
more complex hadronic matrix elements, such as the ones defining the GPDs, 
which are
natural candidates for future studies of cases where the experimental data 
are not numerous enough to allow
for a model independent fitting, and the guidance and intuition of the user is
therefore irreplaceable.
The possibilities of such a method are widely unexplored.

\end{document}